\documentclass[preprint,showpacs,preprintnumbers,nofootinbib]{revtex4}

\usepackage[dvips]{color}
\usepackage{graphicx}
\usepackage{bbm}

\newcommand{\diag}{{\rm Diag}}
\newcommand{\ev}{{\rm eV}}

\newcommand{\bmx}{\left(\begin{array}}
\newcommand{\emx}{\end{array}\right)}

\begin{document}
\title{Getting at large $\theta_{13}^{}$ with almost maximal $\theta_{23}^{}$ from tri-bimaximal mixing}
\author{
Takeshi Araki\footnote{araki@ihep.ac.cn} } 
\affiliation{
Institute of High Energy Physics, Chinese Academy of
Sciences, Beijing 100049, China }

\begin{abstract}
We introduce a small correction term, 
$\delta M_\nu^{}$, in the neutrino sector and examine 
whether a large $\theta_{13}^{}$ and an almost maximal 
$\theta_{23}^{}$ can simultaneously be obtained 
starting from the tri-bimaximal neutrino mixing.
It is found that one can easily gain 
$\theta_{13}^{}\simeq 10^\circ$, which is favored by 
the recent T2K experiment, by taking 
account of the enhancement 
due to the degeneracy among three neutrino masses.
We also find that 
$(\delta M_\nu^{})_{22}^{}=(\delta M_\nu^{})_{33}^{}$ 
is a key condition for $\theta_{23}^{}\simeq 45^\circ$.
\end{abstract}

\maketitle

Thanks to neutrino oscillation experiments \cite{pdg}, 
we currently have convincing evidence that neutrinos 
have tiny masses and mix with each other; the 
latest global analysis of neutrino oscillation data 
yields the following best-fit values with $1\sigma$ 
errors for the oscillation parameters \cite{gfit}:
\begin{eqnarray}
&&\Delta m_{21}^2 = 
(7.59^{+0.20}_{-0.18})\times 10^{-5}_{}~~\ev^2 ,
\nonumber \\
&&\Delta m_{31}^2 =
\left\{\begin{array}{ll}
+(2.45\pm 0.09)\times 10^{-3}_{}~~\ev^2 & 
~~{\rm for~normal~hierarchy~(NH)} \\
-(2.34^{+0.10}_{-0.09}) \times 10^{-3}_{}~~\ev^2 & 
~~{\rm for~inverted~hierarchy~(IH)}
\end{array}\right. ,\nonumber \\
&&\theta_{12}^{} = (34.0 \pm 1.0)^\circ,
~~~~
\theta_{23}^{} = 
\left\{\begin{array}{l}
(45.6^{+3.4}_{-3.5})^\circ \\
(46.1^{+3.5}_{-3.4})^\circ
\end{array}\right. ,
~~~~
\theta_{13}^{} = 
\left\{\begin{array}{l}
(5.7^{+2.2}_{-2.1})^\circ \\
(6.5^{+2.0}_{-2.1})^\circ
\end{array}\right. , \label{eq:exp}
\end{eqnarray}
where upper (lower) values of $\Delta m_{31}^2$, 
$\theta_{23}^{}$ and $\theta_{13}^{}$ correspond to 
the normal (inverted) neutrino mass hierarchy case.
On the theoretical side, it has been known that 
the observed mixing angles can well be described 
by the so-called tri-bimaximal (TB) neutrino mixing 
ansatz \cite{TB}, and a lot of theoretical 
models of the TB mixing have been proposed 
\cite{A4,S3,S4}.

Recently, however, the T2K Collaboration has 
reported possible indications of the 
$\nu_\mu^{}\rightarrow\nu_e^{}$ appearance 
with the following ranges of 
$\theta_{13}^{}$: 
\begin{eqnarray}
&&5.0^\circ < \theta_{13}^{} < 16.0^\circ ~~{\rm for~NH},\\
&&5.8^\circ < \theta_{13}^{} < 17.8^\circ ~~{\rm for~IH}
\end{eqnarray}
at the $90\%$ confidence level \cite{t2k}.
Moreover, the best-fit value of $\theta_{13}^{}$ is 
found to be $\theta_{13}^{}\simeq 9.7^\circ$ for NH 
and $\theta_{13}^{}\simeq 11.0^\circ$ for IH.
Although a magnitude of the confidence level 
is still poor\footnote{
Soon after the T2K report, the MINOS Collaboration has also released new data which indicate $\theta_{13}\neq 0^\circ$ at the $1.5\sigma$ level \cite{MINOS}. Furthermore, evidence for  $\theta_{13}\neq 0^\circ$ at the $>3\sigma$ level has been obtained in a global analysis \cite{fogli}.}, 
if such a large $\theta_{13}^{}$ is 
finally confirmed, 
the TB mixing ansatz would run into trouble
because it predicts $\theta_{13}^{TB}=0^\circ$.
One possible way to save the TB mixing is to 
introduce a small correction term, so that 
one can bring up $\theta_{13}^{}$ to around $10^\circ$.
In this paper, we study whether 
$\theta_{13}^{}\simeq 10^\circ$ can really be 
obtained starting from the TB mixing while maintaining 
the other oscillation parameters within the 
$1\sigma$ ranges of Eq. (\ref{eq:exp}).

We suppose that neutrinos are Majorana particles and
define the zeroth-order charged lepton and neutrino 
mass matrices as
\begin{eqnarray}
M_\ell^0=\diag(m_e^0,m_\mu^0,m_\tau^0), 
~~~M_\nu^0 = V_{}^0 ~\diag(m_1^0 e^{i\rho}_{}, m_2^0 e^{i\sigma}_{}, m_3^0 e^{i\xi}_{})~ (V_{}^0)^T_{},
\label{eq:M0}
\end{eqnarray}
where $m_{e,\mu,\tau}^0$ and $m_{1,2,3}^0$ are 
real and positive parameters, while $\rho$, $\sigma$, 
and $\xi$ are Majorana CP-violating phases. 
$V^0_{}$ is the zeroth-order mixing matrix given by
\begin{eqnarray}
V^0_{} =
\bmx{ccc}
 c_{12}^0 & s_{12}^0 & 0 \\
-\frac{s_{12}^0}{\sqrt{2}} & \frac{c_{12}^0}{\sqrt{2}} & -\frac{1}{\sqrt{2}} \\
-\frac{s_{12}^0}{\sqrt{2}} & \frac{c_{12}^0}{\sqrt{2}} &  \frac{1}{\sqrt{2}}
\emx , \label{eq:V0}
\end{eqnarray}
where $s_{ij}^0=\sin\theta_{ij}^0$, and $\theta_{ij}^0$ 
stands for the zeroth-order mixing angle.
Notice that we have left $\theta_{12}^0$ arbitrary to 
make the discussions as general as possible; thus, 
the following discussions may be applicable for other 
mixing patterns.
For this setup, we introduce small correction terms 
for the charged lepton and neutrino sectors, respectively:
\begin{eqnarray}
M_\ell^{}=M_\ell^0 + \delta M_\ell^{}, ~~~
M_\nu^{}=M_\nu^0+\delta M_\nu^{}. \label{eq:M}
\end{eqnarray}
As we shall explain later, we will consider 
strong enhancement of the correction in the 
neutrino sector.
Hence, we assume that contributions from the 
neutrino sector are dominant and ignore 
$\delta M_\ell^{}$ in the following discussions.

By regarding $\delta M_\nu^{}$ as a small 
perturbation and doing some perturbative calculations, 
the corrected mixing angles are calculated as 
\begin{eqnarray}
&&\sin\theta_{13}^{} =
\left|
\frac{c_{12}^0 P_{13}^{}}{m_1^0 e_{}^{i\rho}-m_3^0} 
+ \frac{s_{12}^0 P_{23}^{}}{m_2^0 e_{}^{i\sigma}-m_3^0}
\right| , \label{eq:mod13}\\
&&\tan\theta_{23}^{} = 
\left| 
1-2
\left[
\frac{s_{12}^0 P_{13}^{}}{m_1^0 e^{i\rho}_{} - m_3^0} - \frac{c_{12}^0 P_{23}^{}}{m_2^0 e^{i\sigma}_{} - m_3^0}
\right]
\right|, \label{eq:mod23}\\
&&\tan\theta_{12} =
t_{12}^0
\left|
1-\frac{1}{s_{12}^0 c_{12}^0} \frac{P_{12}^{}}{m_1^0 e^{i\rho}_{} - m_2^0 e^{i\sigma}_{}}
\right| , \label{eq:mod12}
\end{eqnarray}
where $P_{ij}^{}$ is the element of 
$P=(V^0_{})^T_{} \delta M_\nu^{} V^0_{}=P^T_{}$, 
and we have chosen the basis of $\xi=0$.

Let us roughly estimate the magnitude of 
$\delta M_\nu^{}$ in the case of 
$\theta_{13}^{}\simeq 10^\circ$.
If we suppose the hierarchical neutrino mass 
spectrum and ignore the mixing angles, CP-violating 
phases and order-one coefficients, we obtain 
\begin{eqnarray}
\sin\theta_{13}^{} \simeq 
\frac{\delta M_\nu^{}}{\sqrt{|\Delta m_{31}^2}|}.
\end{eqnarray}
Thus, in order to realize $\theta_{13}^{}\simeq 10^\circ$, 
we need a relatively large $\delta M_\nu^{}$, 
such as $\delta M_\nu^{} \simeq 0.01~\ev$ for 
$M_\nu^0 \simeq 0.05~\ev$.
On the other hand, we can make use of the enhancement 
due to the degeneracy among three neutrino masses, 
as found in 
the studies of radiative corrections \cite{rge,agx}.
For instance, by assuming $m_3^{} = 0.2~\ev$ for 
$\Delta m_{31}^2 = 2.45\times 10^{-3}_{}~\ev^2$, 
we obtain 
\begin{eqnarray}
\sin\theta_{13}^{} \simeq 
\frac{\delta M_\nu^{}}{0.0062~\ev},
\end{eqnarray}
which results in $\delta M_\nu^{} \simeq 0.001~\ev$ 
for $M_\nu^0 \simeq 0.2~\ev$. 
In this way, we can easily generate a relatively large 
$\theta_{13}^{}$ with a naturally small correction 
term.
Of course, we do not have any criterion for 
{\it naturalness}.
Nevertheless, we only focus on the degenerate case 
in this paper. 
More general and systematic studies, including the 
contributions from the charged lepton sector, as 
well as the hierarchical spectrum case 
can be found in Ref. \cite{rod}.

As one can see from Eqs. (\ref{eq:mod13}) and 
(\ref{eq:mod23}), $\theta_{13}^{}$ and 
$\theta_{23}^{}$ are described by the same 
parameters.
Therefore, in general, such a large $\theta_{13}^{}$ 
gives rise to a large deviation of 
$\theta_{23}^{}$ from $45^\circ$ at the same time, 
whereas $\theta_{23}^{}\simeq 45^\circ$ 
is favored by experiments.
This fact might suggest a failure of the TB mixing, 
and we may have to consider a different type 
of mixing, as studied in Refs. \cite{dc,hz,Chao,tetra}.
Meanwhile, if we still believe the TB mixing, it may 
provide us with crucial hints for understanding 
the nature of $\delta M_\nu^{}$.
In other words, the constraint 
$|\theta_{23}^{}-45^\circ|\ll 1$ 
could be reduced to a special condition on 
$\delta M_\nu^{}$.
In fact, some indications along this direction 
are obtained in specific flavor models \cite{model}.
Here, we would like to show this really happens, without 
assuming any flavor model, 
if we impose $\rho=\sigma$\footnote{
This may indicate the exact degeneracy between the 
first and the second eigenvalues of $M_\nu^0$ 
and it could be ensured by a flavor symmetry. 
Or, one can simply presume CP invariance in 
the neutrino sector.
}
for the Majorana phases.
In this case, we can approximately rewrite 
Eqs. (\ref{eq:mod13}) and (\ref{eq:mod23}) as 
\begin{eqnarray}
&&\sin\theta_{13}^{} \simeq
\left|
\frac{1}{m_1^{} e_{}^{i\rho}-m_3^{}}
\left[
c_{12}^0 P_{13}^{} + s_{12}^0 P_{23}^{} 
\right]
\right|,\label{eq:mod13-2}\\
&&\tan\theta_{23}^{} \simeq
\left| 
1-\frac{2}{m_1^{}e^{i\rho}_{} - m_3^{}}
\left[
s_{12}^0P_{13}^{} - c_{12}^0 P_{23}^{}
\right]
\right|\label{eq:mod23-2}
\end{eqnarray}
because we are now considering the 
quasidegenerate neutrino mass spectrum 
and a very small correction term.
(That is, $m_i^0 \simeq m_i^{}$ and 
$m_1^{} \simeq m_2^{}$.)
We next define $\delta M_\nu^{}$ as the most 
general complex symmetric matrix:
\begin{eqnarray}
\delta M_\nu^{}=
\bmx{ccc}
 a & d & e \\
 d & b & f \\
 e & f & c
\emx \label{eq:dMorg}
\end{eqnarray}
and calculate the elements of 
$P=(V^0_{})^T_{} \delta M_\nu^{} V^0_{}$ 
with Eq. (\ref{eq:V0}):
\begin{eqnarray}
&&P_{11}^{}=
(c_{12}^0)^2_{}a + \frac{(s_{12}^0)^2_{}}{2}(b+c+2f)
-\sqrt{2}s_{12}^0 c_{12}^0(d+e) ,
\nonumber \\
&&P_{22}^{}=
(s_{12}^0)^2_{}a + \frac{(c_{12}^0)^2_{}}{2}(b+c+2f)
+\sqrt{2}s_{12}^0 c_{12}^0(d+e) ,
\nonumber \\
&&P_{33}^{}=
\frac{1}{2}(b+c-2f),
\nonumber \\
&&P_{12}^{}=
\frac{s_{12}^0 c_{12}^0}{2}(2a-b-c-2f)
+\frac{(c_{12}^0)^2_{}-(s_{12}^0)^2_{}}{\sqrt{2}}(d+e),
\nonumber \\
&&P_{13}^{}=
\frac{s_{12}^0}{2}(b-c)-\frac{c_{12}^0}{\sqrt{2}}(d-e),
\nonumber \\
&&P_{23}^{}=
-\frac{c_{12}^0}{2}(b-c)-\frac{s_{12}^0}{\sqrt{2}}(d-e).
\end{eqnarray}
By substituting $P_{13}^{}$ and $P_{23}^{}$ into 
Eqs. (\ref{eq:mod13-2}) and (\ref{eq:mod23-2}), 
we finally arrive at surprisingly simple formulas 
\begin{eqnarray}
&&\sin\theta_{13}^{} \simeq \frac{1}{\sqrt{2}}
\left|
\frac{d-e}{m_1^{} e_{}^{i\rho}-m_3^{}}
\right|,\\
&&\tan\theta_{23}^{} \simeq
\left| 
1-\frac{b-c}{m_1^{}e^{i\rho}_{} - m_3^{}}
\right|.
\end{eqnarray}
Then, one can immediately read off that 
$\theta_{23}^{} \simeq 45^\circ$ can be realized by 
requiring $b=c$, and $\theta_{13}^{}\neq 0^\circ$ 
demands $d\neq e$.
We note that these are not the necessary and sufficient 
conditions for $\theta_{23}^{} \simeq 45^\circ$ and 
$\theta_{13}^{}\neq 0^\circ$.
One can find another solution in Ref. \cite{TBR}.

\begin{figure}[ht]
\begin{center}
\includegraphics*[width=0.6\textwidth]{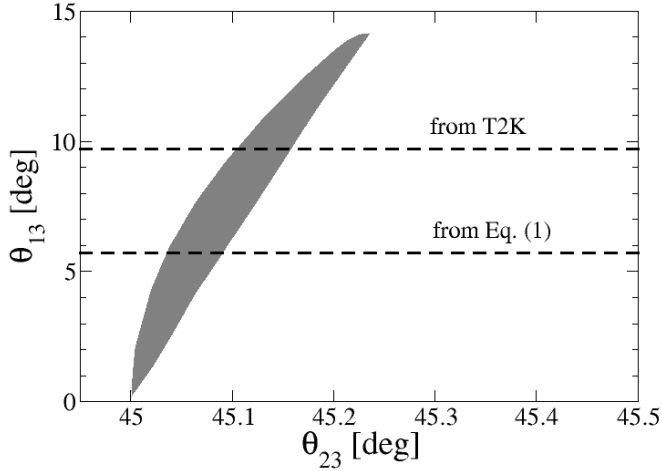}
\caption{\footnotesize $\theta_{13}^{}$ as a function of 
$\theta_{23}^{}$ for the normal mass hierarchy case with 
$m_3^{}=0.2~\ev$. The best-fit values of $\theta_{13}^{}$ 
from the T2K experiment ($9.7^\circ$) and 
Eq. (\ref{eq:exp}) ($5.7^\circ$) are also 
shown as dashed lines.
} \label{fig:a-r}
\end{center}
\end{figure}
To illustrate the consequence gained above, we here 
invent the following simple form of $\delta M_\nu^{}$:
\begin{eqnarray}
\delta M_\nu^{} =
\bmx{ccc}
 0 & |d| & -|e| \\
 |d| & |b|e^{i\psi}_{} & 0 \\
 -|e| & 0 & |b|e^{i\psi}_{}
\emx \label{eq:dM}
\end{eqnarray}
and numerically diagonalize the full mass matrix 
$M_\nu^{}=M_\nu^0 + \delta M_\nu^{}$.
The zeroth-order mass matrix, $M_\nu^0$, is defined 
in Eq. (\ref{eq:M0}), and we fix $\theta_{12}^0$ as 
the TB value, i.e., $s_{12}^0 = 1/\sqrt{3}$.
We vary $|b|, |d|$, and $|e|$ within 
$0$ to $0.01~\ev$, the CP phases ($\rho$ and $\psi$) 
within $0$ to $2\pi$, and $m_{1,2,3}^0$ to be 
consistent with the $1\sigma$ ranges of 
$\Delta m_{31}^2$, $\Delta m_{21}^2$, and 
$\theta_{12}^{}$.
In Fig. \ref{fig:a-r}, we plot $\theta_{13}^{}$ 
as a function of $\theta_{23}^{}$ for the normal 
hierarchy case with $m_3^{}=0.2~\ev$.
As one can see, indeed, we can achieve 
$\theta_{13}^{}\simeq 10^\circ$, which is favored 
by the T2K experiment, while keeping  
$\theta_{23}^{}$ around $45^\circ$.
Several comments are in order.
(1) We have turned off $a$, $f$, and some CP phases 
in Eq. (\ref{eq:dM}) to reduce the number of 
controllable parameters and to minimize the 
calculation.
Even if these parameters are taken into account, 
$\theta_{23}^{}$ stays almost maximal 
due to the condition $b=c$.
(2) If we put a minus sign in front of $\delta M_\nu^{}$, 
$\theta_{23}^{}$ turns out to be less than $45^\circ$.

Lastly, we would like to remark on the possible 
origin of $b=c$. 
In the Zee model \cite{zee}, which is a one-loop 
neutrino mass generation model, the diagonal elements 
of a neutrino mass matrix are vanishing because of 
the Fermi statistics.
Thus, if we implement the Zee mechanism as a finite 
quantum correction \cite{agx,fqc}, $\delta M_\nu^{}$ 
automatically possesses $b=c=0$. 
The result of numerical analysis, in this case, 
is very similar to Fig. \ref{fig:a-r}.

In summary, we have introduced a small correction 
term $\delta M_\nu^{}$ for the tri-bimaximal neutrino 
mixing and examined whether a large 
$\theta_{13}^{}$ and an almost maximal 
$\theta_{23}^{}$ can simultaneously be obtained.
We have found that $\theta_{13}^{}\simeq 10^\circ$, 
which is favored by the T2K experiment, can easily be 
gained by considering the enhancement 
due to the degeneracy among three neutrino masses.
On the other hand, the almost maximal $\theta_{23}^{}$ 
seems to indicate a special condition on 
$\delta M_\nu^{}$, and we have found that 
$(\delta M_\nu^{})_{22}^{}=(\delta M_\nu^{})_{33}^{}$ 
is a key condition for $\theta_{23}^{}\simeq 45^\circ$.
We have also given a simple example of $\delta M_\nu^{}$ 
and shown that $\theta_{13}^{}$ can really be around 
$10^\circ_{}$ while keeping $\theta_{23}^{}\simeq 45^\circ$.

\begin{acknowledgments}
I would like to thank Z. Z. Xing for useful discussions.
This work  was supported in part by the National
Natural Science Foundation of China under Grant 
No. 10875131. 
\end{acknowledgments}

%



\begin{thebibliography}{3}
\bibitem{pdg}
K. Nakamura {\it et al.} (Particle Data Group), 
J. Phys. G {\bf 37}, 075021 (2010).

\bibitem{gfit}
T. Schwetz, M. Tortola, and J. W. F. Valle,
New J. Phys. {\bf 13}, 063004 (2011).

\bibitem{TB}
P. F. Harrison, D. H. Perkins, and W. G. Scott, 
Phys. Lett. B {\bf 530}, 167 (2002); 
Z. Z. Xing, Phys. Lett. B {\bf 533}, 85 (2002); 
P. F. Harrison and W. G. Scott, 
Phys. Lett. B {\bf 535}, 163 (2002).

\bibitem{A4}
E. Ma, 
Phys. Rev. D {\bf 70}, 031901 (2004);
Phys. Rev. D {\bf 73}, 057304 (2006);
G. Altarelli and F. Feruglio, 
Nucl. Phys. B {\bf 720}, 64 (2005);
Nucl. Phys. B {\bf 741}, 215 (2006);
K. S. Babu and X. G. He, 
arXiv:hep-ph/0507217.

\bibitem{S3}
R. Jora, S. Nasri, and J. Schechter,
Int. J. Mod. Phys. A {\bf 21}, 5875 (2006);
C. Y. Chen and L. Wolfenstein, 
Phys. Rev. D {\bf 77}, 093009 (2008);
R. Jora, J. Schechter, and M. N. Shahid, 
Phys. Rev. D {\bf 80}, 093007 (2009) 
[Erratum-ibid. D {\bf 82}, 079902 (2010)].

\bibitem{S4}
C. S. Lam, Phys. 
Lett. B {\bf 656}, 193 (2007); 
Phys. Rev. Lett. {\bf 101}, 121602 (2008); 
Phys. Rev. D {\bf 78}, 073015 (2008); 
arXiv:0907.2206 [hep-ph]; 
Phys. Rev. D {\bf 83}, 113002 (2011).

\bibitem{t2k}
The T2K collaboration, K. Abe {\it et al.}, 
arXiv:1106.2822 [hep-ex].

\bibitem{MINOS}
See the website of the MINOS collaboration, 
http://www-numi.fnal.gov/pr plots/.

\bibitem{fogli}
G. L. Fogli, E. Lisi, A. Marrone, 
A. Palazzo, and A. M. Rotunno,
arXiv:1106.6028 [hep-ph].

\bibitem{rge} 
See, e.g., J. A. Casas {\it et al.},
Nucl. Phys. B {\bf 573}, 652 (2000);
S. Antusch {\it et al.},
Nucl. Phys. B {\bf 674}, 401 (2003).

\bibitem{agx}
T. Araki, C. Q. Geng, and Z. Z. Xing, 
Phys. Lett. B {\bf 699}, 276 (2011). 

\bibitem{rod}
S. Goswami, S. T. Petcov, S. Ray, and W. Rodejohann,
Phys. Rev. D {\bf 80}, 053013 (2009). 

\bibitem{dc}
Z. Z. Xing,
Phys. Lett. B {\bf 696}, 232 (2011);
arXiv:1106.3244 [hep-ph];
S. Zhou, arXiv:1106.4808 [hep-ph].

\bibitem{hz}
X. G. He and A. Zee,
arXiv:1106.4359 [hep-ph].

\bibitem{Chao}
W. Chao and Y. J. Zheng, 
arXiv:1107.0738 [hep-ph].

\bibitem{tetra}
H. Zhang and S. Zhou,
arXiv:1107.1097 [hep-ph].

\bibitem{model}
Y. Shimizu, M. Tanimoto, and A. Watanabe, 
Prog. Theor. Phys. {\bf 126}, 81 (2011); 
E. Ma and D. Wegman, 
arXiv:1106.4269 [hep-ph];
N. Haba and R. Takahashi,
arXiv:1106.5926 [hep-ph].

\bibitem{TBR}
S. F. King,
Phys. Lett. B {\bf 675}, 347 (2009);
S. Morisi, K. M. Patel, and E. Peinado, 
arXiv:1107.0696 [hep-ph].

\bibitem{zee}
A. Zee, Phys. Lett. B {\bf 93}, 389 (1980); 
L. Wolfenstein, Nucl. Phys. B {\bf 175}, 93 (1980). 

\bibitem{fqc}
See, e.g.,
T. Araki, arXiv:1104.1689 [hep-ph]; 
D. A. Sierra and C. E. Yaguna, 
arXiv:1106.3587 [hep-ph].

\end{thebibliography}
\end{document}